\documentclass{rmf-d}
\usepackage{nopageno,rmfbib,multicol,times,epsf,amsmath,amssymb,cite}
\usepackage[latin1]{inputenc}
\usepackage[]{caption2}
\usepackage{graphics}
\def\rmfcornisa{Research OR Education of Physics \hfill\rmf\ {\bf ??} (*?*) ???--???
\hfill MES? A\~NO?}

\def\rmfcintilla{{\it Rev.\ Mex.\ Fis.\/} {\bf ??} (*?*) (????) ???--???}
\clearpage \rmfcaptionstyle 
\begin{document}
\title{   A new theoretical determination of $R_{\tau/P}$ $(P=\pi,K)$  \footnote{Talk given at the 19th International Conference on Hadron Spectroscopy and Structure in memoriam Simon Eidelman (HADRON2021), 26-31 July  (2021), Mexico City (Mexico).} \vspace{-6pt} }

\author{ Ignasi Rosell    }
\address{  Departamento de Matem\'aticas, F\'\i sica y Ciencias Tecnol\' ogicas, Universidad Cardenal Herrera-CEU, \\ CEU Universities, 46115 Alfara del Patriarca, Val\`encia, Spain }
\author{ }
\address{ }
\author{ }
\address{ }
\author{ }
\address{ }
\author{ }
\address{ }
\author{ }
\address{ }
\maketitle
\recibido{day month year}{day month year
\vspace{-12pt}}
\begin{abstract}
\vspace{1em} We have determined $R_{\tau/P}\equiv \Gamma(\tau \to P \nu_\tau [\gamma]) / \Gamma(P \to\mu \nu_\mu[\gamma])$ ($P=\pi, K$). Whereas $P$ decays are calculated by using Chiral Perturbation Theory (ChPT), $\tau$ decays have been studied with an effective approach where ChPT is enlarged by including the lightest resonances and following the high-energy behavior dictated by QCD. These ratios have allowed us to test the lepton universality and the CKM unitarity and also to search for bounds on non-standard interactions. Our results, $\delta R_{\tau/\pi}=(0.18\pm 0.57 )\%$  and $\delta R_{\tau/K}=(0.97\pm 0.58 )\%$, are consistent with the previous theoretical determinations, but with much more robust assumptions, yielding a reliable uncertainty.   \vspace{1em}
\end{abstract}
\keys{   Radiative decays, tau, pion, kaon, lepton universality, effective field theories, CKM unitarity, non-standard interactions. \vspace{-4pt}}
\pacs{   \bf{\textit{12.15.Lk, 12.15.Hh, 12.39.Fe, 12.60.-i, 13.20.Cz, 13.20.Eb, 13.35.Dx, 13.40.Ks.}}    \vspace{-4pt}}
\begin{multicols}{2}

\section{Introduction}

Lepton universality (LU) is an important ingredient of the Standard Model (SM) and a few anomalies have been observed in semileptonic $B$ meson decays. Therefore, searching for other precise tests of LU is very interesting and this is the main reason why we have addressed the calculation of the ratio ($P=\pi,K$)~\cite{Marciano:1993sh, DF,Arroyo-Urena:2021nil}
\begin{equation}\label{eq:MainDef}
 R_{\tau/P}\equiv \frac{\Gamma(\tau \to P \nu_\tau  [\gamma])}{\Gamma(P \to\mu \nu_\mu[\gamma])}=\bigg|\frac{g_\tau}{g_\mu}\Bigg|^2_P R_{\tau/P}^{(0)} \left(1+\delta R_{\tau/P}\right)\,,   
\end{equation}
being $g_\mu=g_\tau$ according to LU, the radiative corrections are parametrized in $\delta R_{\tau/P}$ and $R_{\tau/P}^{(0)}$ is the leading-order result,
\begin{equation}\label{eq:LO}
R_{\tau/P}^{(0)} = \frac{1}{2} \frac{M_\tau^3}{m_\mu^2m_P}\frac{(1-m_P^2/M_\tau^2)^2}{(1-m_\mu^2/m_P^2)^2}\, ,
\end{equation}
which is free from hadronic couplings and CKM matrix elements. Note that a precise test of LU can be done at low energies, where the experiments and the theoretical determinations allow us a comparison with high precision. 

More than twenty-five years ago, $\delta R_{\tau/P}$ was determined in Ref.~\cite{DF}, $\delta R_{\tau/\pi}=(0.16\pm0.14)\%$ and $\delta R_{\tau/K}=(0.90\pm0.22)\%$, and we think that there were important motivations in order to analysis this observable again. 

First of all, different values of $\left|g_\tau/g_\mu\right|$ are reported in the literature depending on the process at hand:
\begin{enumerate}
\item $\Gamma(\tau \to P \nu_\tau  [\gamma]) / \Gamma(P \to\mu \nu_\mu[\gamma])$ ($P=\pi,K$). By using $\delta R_{\tau/P}$ from \cite{DF}, the last HFLAV analysis quoted $\left|g_\tau/g_\mu\right|_\pi=0.9958\pm0.0026$ and $\left|g_\tau/g_\mu\right|_K=0.9879\pm0.0063$, at $1.6\sigma$ and $1.9\sigma$ of LU~\cite{Amhis:2019ckw}.
\item $\Gamma(\tau \to e \bar{\nu}_e \nu_\tau[\gamma])/\Gamma(\mu \to e \bar{\nu}_e \nu_\mu[\gamma])$. This pure leptonic ratio implies $\left|g_\tau/g_\mu\right|=1.0010\pm0.0014$ \cite{Amhis:2019ckw}, at $0.7\sigma$ of LU.
\item $\Gamma(W \to\tau \nu_\tau)/\Gamma(W \to\mu \nu_\mu)$. The weighted average of this $W$-boson branching ratio reads $\left|g_\tau/g_\mu\right|=0.995\pm 0.006$~\cite{Aad:2020ayz, CMS:2021qxj}, at $0.8\sigma$ of LU.
\end{enumerate}
Thus, a new analysis of radiative corrections in $\delta R_{\tau/P}$ was timely to solve these phenomenological discrepancies.

Secondly, there were important theoretical reasons to update the analysis of Ref.~\cite{DF}: different hadronic form factors were considered for virtual- and real-photon correction, they did not satisfy the correct QCD high-energy behavior, violated analyticity, unitarity and the chiral limit at leading non-trivial orders, and a cutoff was used in order to regulate the loop integrals, splitting unphysically long- and short-distance corrections. Moreover, in our view the uncertainties were underestimated in Ref.~\cite{DF}, since they are of the order of an expected purely $\mathcal{O}\left(e^2p^2\right)$ Chiral Perturbation Theory (ChPT) result, {\it i.e.}, a calculation which is not able to include the $\tau$ lepton.

And finally, a new determination of $R_{\tau/P}$ has also allowed us to revisit the CKM unitarity test via $\left|V_{us}/V_{ud}\right|$ in $\Gamma(\tau \to K \nu_\tau[\gamma])/\Gamma(\tau \to \pi \nu_\tau[\gamma])$ or via $\left|V_{us}\right|$ in $\Gamma(\tau \to K \nu_\tau[\gamma])$~\cite{Pich:2013lsa} and to update the constraints on new physics affecting this ratio~\cite{EFTpions, EFTtaudecays1,EFTtaudecays2}.

\section{The calculation}

\begin{figure*}
\begin{center}
\includegraphics{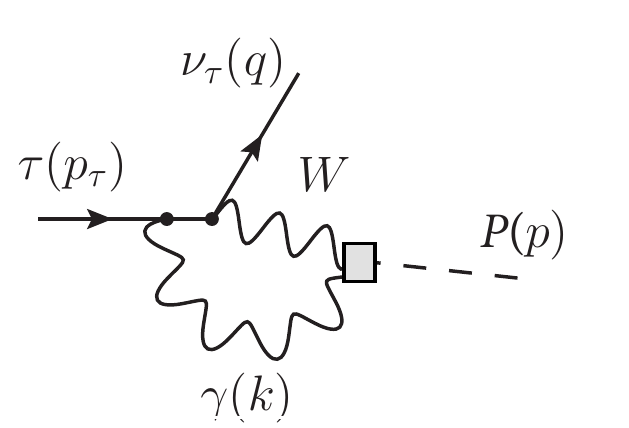} 
\end{center}
\caption{Feynman diagram corresponding to the structure dependent contributions to $\tau \to P \nu_\tau$ decays. The gray shaded box stands for the form factors. \label{feynman}}
\end{figure*}

From a theoretical point of view the calculation of $P$ and $\tau$ decay rates are quite different due to the so different masses: 
\begin{enumerate}
\item Chiral Perturbation Theory (ChPT)~\cite{ChPT}, the very-low effective field theory of QCD to be used for $P$ and light leptons and organized in terms of an increasing number of momenta or masses (the so-called ``chiral'' expansion), allows us to get a model-independent determination of $P$ decays. The only model-dependence appears in the estimation of the counterterms, usually done by matching ChPT with the effective approach at higher energies and including resonances.
\item Resonance Chiral Theory (RChT)~\cite{RChT} offers a well-motivated effective approach to handle QCD at intermediate energies, that is, including resonances, and its expansion parameter is $1/N_C$. One of the main features of this approach is that its low- and high-energy limits are accurate: ChPT results are recovered at low energies and the behavior dictated by QCD is imposed at high energies. In other words, the model-dependence of this approach seems to be under control. Moreover, and following what has been explained previously, RChT is usually used to estimate ChPT counterterms by following a matching procedure.
\end{enumerate}

Despite this difference, the inclusive $P \to\mu \nu_\mu[\gamma]$ and $\tau \to P \nu_\tau  [\gamma]$ decay rates can be parametrized in a similar way~\cite{Marciano:1993sh,CR1,CR2,Arroyo-Urena:2021nil}:
\begin{align}
&\Gamma_{P_{\mu 2 [\gamma]}} = 
\Gamma^{(0)}_{P_{\mu 2 }} \,
S_{ew} \,
\Bigg\{ 1 + \frac{\alpha}{\pi}  \,  F( m_\mu^2/m_P^2)  \Bigg\}
\nonumber \\ 
&\Bigg\{  1 - \frac{\alpha}{\pi}   \Bigg[ \frac{3}{2}  \log \frac{m_\rho}{m_P}   +   c_1^{(P)}  + \frac{m_\mu^2}{m_\rho^2}   \bigg(c_2^{(P)}  \, \log \frac{m_\rho^2}{m_\mu^2}   +  c_3^{(P)} \nonumber \\ &
+ c_4^{(P)} (m_\mu/m_P) \bigg)  -  \frac{m_P^2}{m_\rho^2} \,  \tilde{c}_{2}^{(P)}  \, \log \frac{m_\rho^2}{m_\mu^2}  \Bigg] \Bigg\} \,,
\label{eq:indrate}
\\
&\Gamma_{\tau_{P2[\gamma]}} = 
\Gamma^{(0)}_{\tau_{P2}}\,
S_{ew}\,
\Bigg\{ 1 + \frac{\alpha}{\pi}  \,  G (m_P^2/M_\tau^2)  \Bigg\}
\nonumber \\ 
&\qquad \qquad \Bigg\{  1 - \frac{3\alpha}{2\pi}  \log \frac{m_\rho}{M_\tau} +  \delta_{\tau P}\big|_{\mathrm{rSD}} +  \delta_{\tau P}\big|_{\mathrm{vSD}}\Bigg\} \,,
\label{eq:indratetau}
\end{align}
where in both cases $S_{ew}\simeq  1 + \frac{2 \, \alpha}{\pi}  \log \frac{m_Z}{m_\rho} $ 
is the universal short-distance electroweak correction (canceling in the ratio  $R_{\tau/P}$), the first bracketed terms are the universal long-distance correction (point-like approximation, originally calculated in Ref.~\cite{Kinoshita:1959ha}), the second bracketed terms include the structure-dependent (SD) contributions and $\Gamma^{(0)}_{P_{\mu 2 }}$ and $\Gamma^{(0)}_{\tau_{P2}}$ are the rate in absence of radiative corrections ($F_\pi\sim92$ MeV),
\begin{align}
\Gamma^{(0)}_{P_{\mu 2}} &=\,  \frac{G_F^2 |V_{uD}|^2  F_P^2 }{4 \pi} \, 
m_P  \, m_\mu^2  \, \left(1 - \frac{m_\mu^2}{m_P^2} \right)^2 \,,
\\
\Gamma^{(0)}_{\tau_{P2}}&=\,\frac{G_F^2 |V_{uD}|^2F_P^2 }{8\pi} M_\tau^3\left(1-\frac{m_P^2}{M_\tau^2}\right)^2, 
\label{eq:indratetauLO}
\end{align}
being $D=d,s$ for $P=\pi,K$, respectively. 

For structure-dependent contributions in $P_{\mu 2}$ we have followed the notation proposed in Ref.~\cite{Marciano:1993sh} and we have considered the numerical values for $c_n^{(P)}$ which were reported in Refs.~\cite{CR1,CR2}. On the other hand, structure-dependent contributions in $\tau_{P2}$ have been split into real-photon (rSD) and virtual-photon (vSD) corrections: rSD corrections have been taken from Ref.~\cite{form_factors1} and vSD corrections have been calculated from scratch~\cite{Arroyo-Urena:2021nil} by following the same effective approach used for rSD corrections and for the estimation of the local counterterms in $P \to\mu \nu_\mu[\gamma]$. The Feynman diagram corresponding to this new calculation is shown in Figure~\ref{feynman}.

In the end, all the model-dependence of our result is related to the hadronization of the QCD currents, relevant for the estimation of the counterterms in $P_{\mu 2}$ and for SD contributions in $\tau_{P2}$, and it is encoded in the form factors $F^P_{V,A} (W^2, k^2)$ and $B(k^2)$,
\begin{eqnarray}
F^{P}_V(W^2,k^2)&=&\frac{-N_C M_V^4}{24\pi^2F_P(k^2-M_V^2)(W^2-M_V^2)}\, ,  \nonumber \\
F_{A}^P(W^2,k^2)&=&\frac{F_P}{2} \frac{M_A^2 -2M_V^2 -k^2 }{(M_V^2-k^2)(M_A^2-W^2)}\,, \nonumber \\
B(k^2)&=& \frac{F_P}{M_V^2-k^2}\,, \label{form_factors_scenariob}
\end{eqnarray}
which were taken from Refs.~\cite{form_factors1,form_factors2}. They are found by imposing well-behaved two- and three-point Green functions and considering the chiral and $U(3)$ flavor limits. Note that $M_V$ and $M_A$ stand for the vector and axial-vector resonance masses, $M_V=M_\rho$, $M_A=M_{a_1}$ and $M_V=M_K*$, $M_A\sim M_{f_1}$ for the pion and kaon case, respectively. 

\begin{table*}[t!!!!]
\begin{center}
\renewcommand{\arraystretch}{1.5}
\begin{tabular}{|c|c|c|c|}
\hline
  Contribution & $\delta R_{\tau/\pi}$   & $\delta R_{\tau/K}$ &  Ref.  \\[5pt]
\hline \hline 
SI &  $+1.05\%$& $+1.67\%$ &\cite{DF} \\
rSD  &$+0.15\%$   &$+(0.18\pm 0.05)\%$ & \cite{CR1,form_factors1} \\
vSD & $-(1.02\pm 0.57 )\%$& $-(0.88\pm 0.58)\%$ & new~\cite{Arroyo-Urena:2021nil} \\
\hline \hline
Total & $+(0.18\pm 0.57 )\%$ & $+(0.97\pm 0.58 )\%$ & new~\cite{Arroyo-Urena:2021nil}
\\
\hline
\end{tabular}
\end{center}
\caption{Numerical values of the different photonic contributions to $\delta R_{\tau/P}$: Structure Independent (SI), real-photon Structure Dependent (rSD) and virtual-photon Structure Dependent (vSD). Errors are not reported for contributions where the uncertainties are negligible for the level of accuracy of this analysis, that is, lower than $0.01\%$.}
\label{tab:tab2}
\end{table*}

The technicalities of the calculation can be found in Ref.~\cite{Arroyo-Urena:2021nil} and we show in Table~\ref{tab:tab2} the different contributions. The final result reads~\cite{Arroyo-Urena:2021nil}
\begin{eqnarray} 
\delta R_{\tau/\pi} &=& (0.18\pm 0.57 )\% \,, \nonumber \\
\delta R_{\tau/K}&=& (0.97\pm 0.58 )\%\,. \label{finalresult}
\end{eqnarray} 
The most important source of uncertainty comes from the estimation of the counterterms appearing in vSD contributions of $\tau \to P \nu_\tau$ ($\pm 0.57\%$ and $\pm 0.58\%$ for the pion and kaon case, respectively, see Table~\ref{tab:tab2}) and we have estimated this uncertainty by considering the running of the counterterms between $0.5\,$and $1.0\,$GeV ($\pm 0.52\%$) and comparing also our result with the result obtained with a less general scenario where only well-behaved two-point Green functions and a reduced resonance Lagrangian is used ($\pm 0.22\%$ and $\pm 0.24\%$ for the pion and kaon case, respectively), that is, modifying the form factors of (\ref{form_factors_scenariob}).~\footnote{Let us spotlight the conservative estimation we have followed. Taking into account the different scales, counterterms in RChT are expected to be lower than in ChPT. However, with our estimation the counterterms affecting the vSD corrections in $P_{\mu 2}$ and $\tau_{P2}$ are of similar size.}

As it has been stressed previously, this result is compatible with Ref.~\cite{DF}, $\delta R_{\tau/\pi}=(0.16\pm0.14)\%$ and $\delta R_{\tau/K}=(0.90\pm0.22)\%$. Although their central values agree remarkably, this is only a coincidence, since uncertainties were underestimated in Ref.~\cite{DF}, where they have approximately the size which would be expected in a purely Chiral Perturbation Theory computation. Moreover, let us stress again the inconsistencies in Ref.~\cite{DF}: the hadronization of the QCD currents was different for real- and virtual-photon corrections, did not satisfy the correct QCD high-energy behavior, violated analyticity, unitarity and the chiral limit at leading non-trivial orders, and a cutoff was used in order to regulate the loop integrals, separating for no reason long- and short-distance corrections. 

\section{Applications}

Our result has given different interesting applications or by-products:
\begin{enumerate}
\item Lepton universality test. Considering our final result of (\ref{finalresult}) in (\ref{eq:MainDef}), one is able to assess the LU~\cite{Arroyo-Urena:2021nil},
\begin{align} 
&\left|\frac{g_\tau}{g_\mu}\right|_\pi =0.9964 \pm 0.0028_{\mathrm{th}}\pm 0.0025_{\mathrm{exp}} \qquad \nonumber \\ & \qquad \qquad \qquad \qquad = 0.9964\pm 0.0038\,, \nonumber \\
&\left|\frac{g_\tau}{g_\mu}\right|_K=0.9857\pm  0.0028_{\mathrm{th}}\pm 0.0072_{\mathrm{exp}} \qquad \nonumber \\ & \qquad  \qquad \qquad \qquad =0.9857\pm 0.0078\,, 
\end{align}
at $0.9\sigma$ and $1.8\sigma$ of LU, respectively. We can compare these results with the HFLAV analysis of Ref.~\cite{Amhis:2019ckw}, $\left|g_\tau/g_\mu\right|_\pi=0.9958\pm0.0026$ and $\left|g_\tau/g_\mu\right|_K=0.9879\pm0.0063$ (at $1.6\sigma$ and $1.9\sigma$ of LU), where $\delta R_{\tau/P}$ was taken from Ref.~\cite{DF}. As it can be observed, our results are closer to LU.
\item Radiative corrections of the individual $\tau \to P \nu_\tau[\gamma]$ decays. We can also use our results to calculate the radiative corrections of the individual $\tau_{P2[\gamma]}$ decays, $\Gamma_{\tau_{P2[\gamma]}} =  \Gamma^{(0)}_{\tau_{P2}} S_{\rm ew}\left( 1 +\delta_{\tau P} \right)$, where $\delta_{\tau P}$ includes all SI and SD radiative corrections in (\ref{eq:indratetau}) and we have found~\cite{Arroyo-Urena:2021nil}:
\begin{eqnarray} 
\delta_{\tau \pi} &=& -( 0.24 \pm 0.56 ) \%\, , \nonumber \\
\delta_{\tau K} &=& -(0.15 \pm 0.57) \%\, . \label{delta}
\end{eqnarray}
\item CKM unitarity test via $\left|V_{us}/V_{ud}\right|$. The ratio
\begin{equation}\label{eq:Vusdet}
\frac{\Gamma(\tau\to K\nu_\tau[\gamma])}{\Gamma(\tau\to \pi\nu_\tau[\gamma])}=\frac{|V_{us}|^2F_K^2}{|V_{ud}|^2F_\pi^2}\frac{(1\!-\!m_{K}^2/M_\tau^2)^2}{(1\!-\!m_{\pi}^2/M_\tau^2)^2}\left(1\!+\!\delta\right)\!,
\end{equation}
can be used to extract $\left|V_{us}/V_{ud}\right|$. Taking into account that the radiative correction in (\ref{eq:Vusdet}) can be extracted from (\ref{delta}), $\delta = \delta_{\tau K}- \delta_{\tau \pi}= (0.10\pm0.80)\%$, it is found~\cite{Arroyo-Urena:2021nil}:
\begin{align} 
& \bigg|\frac{V_{us}}{V_{ud}}\bigg|=0.2288 \pm 0.0010_{\mathrm{th}} \pm 0.0017_{\mathrm{exp}} \qquad \nonumber \\ & \qquad \qquad \qquad \qquad = 0.2288\pm0.0020\,, \label{ourVusVud}
\end{align}
which is $2.1\sigma$ away from unitarity. Note that this analysis was not done in Ref.~\cite{Amhis:2019ckw}. Our result is consistent with recent Ref.~\cite{Seng:2021nar}, $|V_{us}/V_{ud}|=0.2291\pm0.0009$, obtained by considering kaon semileptonic decays.
\item CKM unitarity test via $\left|V_{us}\right|$. Alternatively, one can extract $|V_{us}|$ directly from the $\tau \to K \nu_\tau[\gamma]$ decays, $\Gamma_{\tau_{K2[\gamma]}} =  \Gamma^{(0)}_{\tau_{K2}} S_{\rm ew}\left( 1 +\delta_{\tau K} \right)$. Using our value of $\delta_{\tau K}$ of (\ref{delta}), one finds~\cite{Arroyo-Urena:2021nil}: 
\begin{align} \label{ourVus}
 &|V_{us}|=0.2220 \pm  0.0008_{\mathrm{th}}  \pm 0.0016_{\mathrm{exp}} \qquad \nonumber \\ & \qquad \qquad \qquad \qquad = 0.2220\pm 0.0018 \,, 
\end{align}
at $2.6\sigma$ from unitarity. Again we can compare this result with the HFLAV analysis of Ref.~\cite{Amhis:2019ckw}, $ |V_{us}|=0.2234 \pm 0.0005_{\mathrm{th}} \pm 0.0014_{\mathrm{exp}} = 0.2234
\pm 0.0015$, at $1.3\sigma$ from unitarity. The difference is mainly due to the different experimental inputs considered in Ref.~\cite{Amhis:2019ckw} and here. Again (\ref{ourVus}) is compatible with Ref.~\cite{Seng:2021nar}, $|V_{us}|=0.2231\pm0.0006$, obtained using kaon semileptonic decays. Note that our uncertainties in (\ref{ourVusVud}) and (\ref{ourVus}) do not reach the level of uncertainty of Ref.~\cite{Seng:2021nar} because of the limited statistics in $\tau$ decays, so that one would need to improve the related measurements in Belle II~\cite{Belle-II:2018jsg} to get the same precision.

\item Searching for new physics in $\tau \to P \nu_\tau[\gamma]$ decays. Our results can also be used in order to search for non-SM interactions in $\tau \to P \nu_\tau[\gamma]$ decays: 
\begin{equation}
\Gamma(\tau\!\to\! P\nu_\tau[\gamma]) \!=\!  \Gamma^{(0)}_{\tau_{P2}}  \left|\frac{\widetilde{V}_{uD}}{V_{uD}}\right|^2\!\! S_{\rm ew}\! \left( 1 \!+\!\delta_{\tau P}  \!+\! 2 \Delta^{\tau P}\right) \!,
\end{equation}
being $D=d,s$ for $P=\pi,K$, respectively. $\Delta^{\tau P}$ contains the leading-order new-physics corrections not present in $\widetilde{V}_{uD}=(1 + \epsilon^e_L + \epsilon^e_R )V_{uD}$, directly incorporated by considering 
$V_{uD}$ from nuclear $\beta$ decays~\cite{EFTtaudecays1,EFTtaudecays2}:
\begin{equation}
\Delta^{\tau P} = \epsilon^\tau_L-\epsilon^e_L-\epsilon^\tau_R-\epsilon^e_R-\frac{m_P^2}{M_\tau(m_u+m_D)}\epsilon^\tau_P \,.
\end{equation}
Using our estimations of $\delta_{\tau P}$ and $|V_{us}/V_{ud}|$ of (\ref{delta}) and (\ref{ourVusVud}) respectively, we have found~\cite{Arroyo-Urena:2021nil}:
 \begin{eqnarray}
\Delta^{\tau \pi} &=& -(0.15\pm0.72)\%\,, \nonumber \\
\Delta^{\tau K} &=&-(0.36\pm1.18)\%\, ,
\end{eqnarray}
to be compared to $\Delta^{\tau \pi} = -( 0.15 \pm 0.67 ) \%$ in Ref.~\cite{EFTtaudecays1} and $\Delta^{\tau \pi} = -( 0.12 \pm 0.68 ) \%$ and $\Delta^{\tau K} = -(0.41 \pm 0.93) \%$ in Ref.~\cite{EFTtaudecays2}. These values are given at a scale of $\mu=2$ GeV in the $\overline{\mathrm{MS}}$-scheme. All these estimations are consistent with each other and compatible with the SM, $\Delta^{\tau P}=0$.
\end{enumerate}

For all these results masses and branching ratios have been extracted from the PDG~\cite{PDG}, $|V_{ud}|=0.97373\pm0.00031$ from Ref.\cite{Hardy:2020qwl}, $S_{\rm ew}=1.0232$ from Ref.~\cite{Marciano:1993sh} and meson decay constants from the FLAG analysis~\cite{Aoki:2019cca}: $F_K/F_\pi=1.1932\pm 0.0019$, $\sqrt{2} F_\pi=(130.2\pm0.8)~$MeV and $\sqrt{2} F_K=(155.7\pm0.3)~$MeV.

\section{Conclusions}

We have calculated the radiative corrections in the ratio ($P=\pi,K$)
\begin{equation}
 R_{\tau/P}\equiv \frac{\Gamma(\tau \to P \nu_\tau  [\gamma])}{\Gamma(P \to\mu \nu_\mu[\gamma])}= R_{\tau/P}^{(0)} \left(1+\delta R_{\tau/P}\right)\,,   
\end{equation}
where $R_{\tau/P}^{(0)}$ is the leading-order result. We have found $\delta R_{\tau/\pi}=(0.18\pm 0.57 )\%$  and $\delta R_{\tau/K}=(0.97\pm 0.58 )\%$~\cite{Arroyo-Urena:2021nil}, compatible with the result reported more than twenty-five years ago in Ref.~\cite{DF}, $\delta R_{\tau/\pi}=(0.16\pm0.14)\%$ and $\delta R_{\tau/K}=(0.90\pm0.22)\%$. However, our approach has much more robust theoretical assumptions, resulting in a reliable uncertainty. 

Our results have also been used to different timely topics: to test the lepton universality; to extract the CKM elements $\left|V_{us}/V_{ud}\right|$ and $\left|V_{us}\right|$ and, consequently, to test the CKM unitarity; and to bind effective new-physics interactions.

\section*{Acknowledgments}

I wish to thank the organizers for the, despite being online, pleasant conference. I wish to thank also M. A.~Arroyo-Ure\~na, G.~Hern\' andez-Tom\'e, G.~L\'opez-Castro and P.~Roig, since the work presented here has been done with them and also for their helpful comments to prepare these proceedings. This work has been supported in part by the Spanish Government [MCIN/AEI/10.13039/501100011033, Grant No. PID2020-114473GB-I00]; by the Generalitat Valenciana [PROMETEU/2021/071] and by the Universidad Cardenal Herrera-CEU [INDI20/13].



\end{multicols}
\medline
\begin{multicols}{2}
%


\end{multicols}
\end{document}